\documentclass[a4paper,reqno,12pt,draft]{article}
\usepackage{amssymb,euscript,bbold}
\newcommand{\be}{\begin{equation}}
\newcommand{\ee}{\end{equation}}
\newcommand{\ba}{\begin{eqnarray}}
\newcommand{\ea}{\end{eqnarray}}

\newcommand{\ft}{\footnote}

\begin{document}
\input{epsf}

\begin{flushright}
RUNHETC-2001-03
\end{flushright}
\begin{flushright}
\end{flushright}
\begin{center}
\Large{\sc Confining strings from $G_2$-holonomy spacetimes.}\\
\bigskip
{\sc B.S. Acharya}\ft{bacharya@physics.rutgers.edu}\\
\smallskip\large
{\sf Department of Physics,\\
Rutgers University,\\ 126 Freylinghuysen Road,\\ NJ 08854-0849.}

\end{center}
\bigskip
\begin{center}
{\bf {\sc Abstract}}
\end{center}
The low energy physics of $M$ theory near certain singularities of
$G_2$-holonomy spaces can be described by pure ${\cal N}$ = $1$ super
Yang-Mills theory in four dimensions. In this note we consider the cases
when the gauge group is $SO(2n)$, $E_6$, $E_7$ or $E_8$.
Confining strings with precisely the expected charges are
naturally identified in proposed ``gravity duals'' of these
singular $M$ theory spacetimes. 

\newpage


In \cite{vaf} Vafa described a duality between the wrapped D6-brane system on
${T^*}({S^3})$ - ie the {\it deformation} of the conifold singularity in three complex
dimensions - and the closed Type IIA string backround which is the {\it resolution} of
the same conifold singularity plus Ramond-Ramond fluxes.
More precisely the duality relates open string amplitudes on the
D6-brane side to closed string amplitudes on the closed string side.

In \cite{bsa} we showed that both of the Type IIA string backgrounds involved in
this duality arise from $G_2$-holonomy vacua of $M$ theory. In \cite{avm},
it was proposed that the two sides of this story are continuously connected
in the one complex dimensional moduli space - the two topologically distinct
7-manifolds being related by a flop transition.
On the D6-brane
side the corresponding 7-manifold is a singular orbifold of the total space
$S({S^3})$ of the spin bundle over $S^3$ which is well known to admit a 
$G_2$-holonomy metric \cite{brysal}. If the IIA background has $n$ D6-branes then the
singularities are a family of $A_{n-1}$ orbifold singularities in $\mathbb{R^4}$
fibered over the $S^3$ which is the zero section of $S({S^3})$. On the closed
string side the 7-manifold is simply $S({{S^3}/{\mathbb{Z_n}}})$ - the standard
spin bundle over the Lens space ${S^3}/{\mathbb{Z_n}}$ - also a smooth $G_2$-holonomy
manifold.

A natural question to consider is whether or not this kind of duality extends to other
gauge groups? One can easily replace the $S^3$ family of $A_{n-1}$ orbifold singularities
by $D_{k}$, $E_6$, $E_7$ or $E_8$ singularities. The ``gravity duals'' of
$M$ theory on these singular
$G_2$-holonomy spaces would then naturally
be given by $M$ theory on  $S({{S^3}/{\mathbb{\Gamma}}})$,
with $\mathbb{\Gamma}$ the corresponding $D_{k}$ or $E_i$ type subgroup of $SU(2)$.
For the $D_{k}$ cases this has recently been studied in \cite{vafsin} where several
positive checks of the duality were made.

At low energies below the scale set by the $S^3$, the physics of $M$ theory on
a $G_2$-holonomy, $S^3$
family of $A$, $D$ or $E$ singularities is described by ${\cal N}$ = 1 super Yang-Mills
theory in four dimensions with $A$, $D$ or $E$ gauge group. These gauge
theories are expected to confine at low energies. A test of a dual description of these
$M$ theory backgrounds - as $M$ theory on the smooth $G_2$-holonomy spaces
$S({{S^3}/{\mathbb{\Gamma}}})$ - would be to identify the confining strings. That  is
the purpose of this note.

Topologically,  $S({{S^3}/{\mathbb{\Gamma}}})$ is equivalent to $\mathbb{R^4} {\times}
{S^3}/{\mathbb{\Gamma}}$. Strings in four dimensions can be formed by wrapping $M$2-branes
on one-cycles or $M$5-branes on four-cycles.
In this case, the 7-manifold has one-cycles but
no four-cycles, so the natural candidates for confining strings are $M$2-branes wrapping
one-cycles in ${S^3}/{\mathbb{\Gamma}}$.

The charges of the strings obtained by wrapping the $M$2-branes are given by
$H_{1}(S({{S^3}/{\mathbb{\Gamma}}}))$ $\cong$ $H_{1}({S^3}/{\mathbb{\Gamma}})$
$\cong$ $\mathbb{\Gamma}/[\mathbb{\Gamma},\mathbb{\Gamma} ]$. In the second isomorphism
we use the fact that the first homology group of a manifold
is isomorphic to the abelianisation\ft{The
abelianisation of a finite group is its quotient by the group
generated by all commutators. This group also plays a
crucial role in classifying bound states of D-branes wrapping
submanifolds with non-trivial fundamental group \cite{gopvaf}.}
of its fundamental group. The fundamental group of the
manifold under consideration is simply $\mathbb{\Gamma}$.

So, in order to determine the charges of our candidate strings we need to compute the
abelianisations of all of the finite subgroups of $SU(2)$.

For $\mathbb{\Gamma}$ $\cong$ $\mathbb{Z_n}$, the gauge group is locally $SU(n)$.
Since $\mathbb{Z_n}$ is abelian, its commutator subgroup is trivial and
hence the charges of our strings take values in $\mathbb{Z_n}$. Since this is the
center of $SU(n)$ this is the expected answer for a confining $SU(n)$ theory.

For  $\mathbb{\Gamma}$ $\cong$ $\mathbb{D_{k-2}}$, the binary dihedral group of order
$4k-8$, the local gauge group of the Yang-Mills theory is $SO(2k)$. The binary dihedral
group is generated by two elements $\alpha$ and $\beta$ which obey the relations
\be
{\alpha}^2 = {\beta}^{k-2}
\ee
\be
\alpha \beta = {\beta}^{-1} \alpha
\ee
\be
{\alpha}^4 = {\beta}^{2k-4} = 1
\ee

To compute the abelianisation of  $\mathbb{D_{k-2}}$, we simply take these relations and
impose that the commutators are trivial. From the second relation this implies that
\be
\beta = {\beta}^{-1}
\ee
which in turn implies that
\be
{\alpha}^2 = 1 \;\;\;\; for\;\;\;\;\; k = 2p
\ee
and
\be
{\alpha}^2 = \beta \;\;\;\;\; for\;\;\;\;\; k = 2p+1
\ee
Thus, for $k=2p$ we learn that the abelianisation of $\mathbb{D_{k-2}}$ is isomorphic
to $\mathbb{{Z_2}{\times}Z_2}$, whereas for $k=2p+1$ it is isomorphic to $\mathbb{Z_4}$.
These groups are respectively the centers of $Spin(4p)$ and $Spin(4p+2)$. This is the
expected answer for the confining strings in $SO(2k)$ super Yang-Mills which can be coupled
to spinorial charges.

To compute the abelianisations of the binary tetrahedral (denoted ${\cal T}$),
octahedral (${\cal O}$) and icosahedral (${\cal I}$)
groups which correspond respectively to $E_6$, $E_7$ and $E_8$ super Yang-Mills theory,
we utilise the fact that the order of $G/[G,G]$ - with $G$ a finite group -
is the number of
inequivalent one dimensional representations of $G$. The representation theory of the
finite subgroups of $SU(2)$ is described through the Mckay correspondence by the
representation theory of the corresponding Lie algebras. In particular the dimensions
of the irreducible representations of ${\cal T,O}$ and ${\cal I}$ are given by the
coroot integers (or dual Kac labels) of the affine Lie algebrae associated to $E_6$,
$E_7$ or $E_8$ respectively. From this we learn that the respective orders
of ${\cal T}/[{\cal T},{\cal T}]$,  ${\cal O}/[{\cal O},{\cal O}]$ and
${\cal I}/[{\cal I},{\cal I}]$ are three, two and one. Moreover, one can easily
check that ${\cal T}/[{\cal T},{\cal T}]$ and ${\cal O}/[{\cal O},{\cal O}]$ are
$\mathbb{Z_3}$ and $\mathbb{Z_2}$ respectively, by examining their group relations.
Thus we learn that   ${\cal T}/[{\cal T},{\cal T}]$,  ${\cal O}/[{\cal O},{\cal O}]$ and
${\cal I}/[{\cal I},{\cal I}]$ are, respectively isomorphic to the centers
$Z({E_6})$, $Z({E_7})$ and $Z({E_8})$ in perfect agreement with the expectation that
the super Yang-Mills theory confines.

\bigskip
\Large
\noindent
{\bf \sf Acknowledgements.}
\normalsize

We would like to thank G. Moore, C. Vafa and E. Witten for encouraging discussions.


\begin{thebibliography}{5}
\bibitem{vaf} C. Vafa, {\sf Superstrings and Topological Strings at Large n,}
hep-th/0008142.
\bibitem{bsa} B.S. Acharya, {\sf On Realising ${\cal N}$ $=$ $1$ super Yang-Mills in
$M$ theory,} hep-th/0011089.
\bibitem{avm} M. Atiyah, J. Maldacena, and C. Vafa, {\sf An $M$ theory Flop as a Large
n Duality,} hep-th/0011256.
\bibitem{brysal} R. Bryant and S. Salamon, {\sf On the Construction of Some Complete
Metrics With Exceptional Holonomy,} Duke Math Journal {\bf 58} 3 (1989) 829.
\bibitem{vafsin} S. Sinha and C. Vafa, {\sf SO and Sp Chern-Simons at Large n},
hep-th/0012136.
\bibitem{gopvaf} R. Gopakumar and C. Vafa,
{\sf Branes and Fundamental Groups,} ATMP 2 (1998) 399.
hep-th/9712048.
\end{thebibliography}
\end{document}